\documentclass[12pt]{article}%
\usepackage{amsmath}
\usepackage{graphicx}%
\usepackage{amsfonts}%
\usepackage{amssymb}

\begin{document}

\author{C. S. Unnikrishnan\thanks{E-mail address: unni@tifr.res.in}\\\textit{Gravitation Group, Tata Institute of Fundamental Research,}\\\textit{Homi Bhabha Road, Mumbai - 400 005, India}}
\title{Cosmic Relativity: The Fundamental Theory of Relativity, its Implications, and
Experimental Tests}
\maketitle

\begin{abstract}
In this paper I argue for a reassessment of special relativity. The
fundamental theory of relativity applicable in this Universe has to be
consistent with the existence of the massive Universe, and with the effects of
its gravitational interaction on local physics. A reanalysis of the situation
suggests that all relativistic effects that are presently attributed to
kinematics of relative motion in flat space-time are in fact gravitational
effects of the nearly homogeneous and isotropic Universe. The correct theory
of relativity is the one with a preferred cosmic rest frame. Yet, the theory
preserves Lorentz invariance. I outline the new theory of Cosmic Relativity,
and its implications to local physics, especially to physics of clocks and to
quantum physics. This theory is generally applicable to inertial and
noninertial motion. Most significanlty, experimental evidence support and
favour Cosmic Relativity. There are observed effects that can be consistently
explained only within Cosmic Relativity. The most amazing of these is the
dependence of the time dilation of clocks on their `absolute' velocity
relative to the cosmic rest frame. Important effects on quantum systems
include the physical cause of the Thomas precession responsible for part of
the spectral fine structure, and the phase changes responsible for the
spin-statistics connection. At a deeper level it is conlcuded that relativity
in flat space-time with matter reiterates Mach's principle. There will not be
any relativistic effect in an empty Universe.

\end{abstract}

\section{Introduction}

In this introductory section I point out the need for a reassessment of
Special Relativity (SR). Apart from strong theoretical motivations, in the
unavoidable and inseparable presence of a vast and massive Universe, there is
compelling experimental evidence that suggests that SR has to be replaced by a
more fundamental theory of relativity in flat space-time. A calculation of the
gravitational time dilation effects of galaxies and other masses on clocks
moving relative to the CMBR rest frame shows that these effects are as large
as the effects predicted by SR. But there is no room for both the kinematical
SR effects and gravitational effects to be present together since experimental
data at 0.1\% accuracy allow only one of them to be applicable. But galaxies
and the massive Universe \emph{exist} and therefore those gravitational
effects should be real and ever present. This rules out kinematics as the
basis of relativistic effects, and immediately shows the need to replace
special relativity with a new theory of relativity that relies on the
gravitational effects of the Universe.\ Also, there is unambiguous evidence
from experiments that there are measurable physical effects on clocks that
depend on the velocity of the frame in which they are compared, with respect
to a cosmic absolute rest frame determined by the average gravity of the
Universe. \ 

Our Universe, ``once given'' is observed to be homogeneous and isotropic on
large scales. Recently it is also known that the average density of matter and
energy in this Universe is very close to being `critical', making the Universe
spatially flat. Since we also know that the rate of expansion of the Universe
is very slow at present, only about $2\times10^{-18}$ m/s over a length scale
of 1 meter, we can also take the expansion rate as negligible for time scales
and length scales relevant to laboratory experiments. These facts are
represented by the near equality of the Robertson-Walker metric for the
Universe with matter and energy at critical density (the critical Universe),
and the Minkowski metric for empty space-time. In this special situation
space-time is seemingly flat even in a massive Universe in which every
physical system is interacting gravitationally with a large amount of matter.
But this superficial equivalence that makes SR seemingly correct breaks in a
deeper analysis.

The Special theory of Relativity was formulated in the gravity-free empty
space-time and there is a general belief that the theory is correct and well
tested, despite the knowledge that space-time is not empty or free of gravity.
SR rejected absolute space and absolute time, it rejected preferred frames,
and it rejected anything that can act as a preferred frame, like the
electromagnetic ether. In SR all frames are equivalent, and no experiment
within a uniformly moving frame can reveal the velocity of the frame relative
to any other hypothetical absolute frame. More over, all observers in inertial
motion are equivalent and any such observer is entitled to the claim of being
at a `state of rest', with his local clocks always registering the maximum
`aging' compared to any other clock \emph{moving relative} to him.

\emph{But we know for a fact that there is a preferred frame, and also
absolute time}. The frame in which the Cosmic Microwave Background Radiation
(CMBR) has no dipole anisotropy is the absolute rest frame, and the value of
the monotonically changing temperature provides the absolute time
\cite{unni-twin}. No observer has a right to claim that he is at rest even in
inertial motion, because that will be in conflict with his measurement of the
properties of the Universe. How can these facts be consistent with Special Relativity?

Aren't we misled in a serious way? Isn't possible that the near identity of
the space-time of the massive and vast Universe and that of the empty space
lead us to the illusion that the correct theory of relativity is special
relativity in which time and space intervals are modified by mere kinematics
in empty space, with all the experimental tests seemingly matching with the
predictions of SR, whereas all the effects are in fact due to the ever present
gravitational interaction of the entire Universe? One may dismiss this
possibility, in fact too easily, and might argue that even if that was the
case, if there is no way of distinguishing between two cases -- one in which
all the physical effects are due to kinematics as described SR, and another in
which the effects are dynamical due to the gravity of the Universe -- then one
may chose either as valid. Even admitting this much is a large change in our
understanding of relativity, since the possibility of an equally valid new
theory emerges. But, in fact, the choice does not exist because the massive
Universe exists in which everything is gravitationally linked to everything
else. Comparison with experimental evidence and also requirement of general
applicability and consistency favour the new theory as the more correct theory
of relativity in flat space-time, as we shall see.

It turns out that there are large differences between the two cases. In SR all
inertial frames are equivalent. In SR the physical effects are the same
irrespective of whether there is any matter in the Universe or not. In the new
theory that acknowledges the presence and gravitational effect of the cosmos,
which I call \emph{Cosmic Relativity}, there is exactly one preferred frame,
and all other frames are in relative motion with respect to the preferred
frame. The preferred unique frame is the one that is at rest with respect to
the average matter distribution, and operationally this frame is easily
identified as the one in which there is no dipole anisotropy of the CMBR. The
monotonically varying average temperature of the CMBR provides the absolute
time for every observer. Yet, there are modifications of time and space
intervals in moving frames due to the gravity of the matter in the Universe
because motion modifies the gravitational potentials. In particular,
\emph{when two clocks are compared in a frame moving with respect to the
cosmic frame, they will show an asymmetry in their time dilation depending on
the velocity of the frame with respect to the cosmos}. A dramatic consequence
is that a clock that is moving in a particular frame can `age' faster than a
clock that is stationary in the same frame, in complete contrast to the SR
prediction where it ages slower. This remarkable prediction of Cosmic
Relativity (CR) has in fact been seen in earlier clock comparison experiments.
Cosmic Relativity is favoured by experimental evidence as the correct theory
of relativity. The importance of a reassessment in the light of these findings
and experimental evidence cannot be overemphasized, especially just before the
100th successful year of SR.

In the rest of the paper I outline the details of the new theory of Cosmic
Relativity, its predictions, the differences with SR, evidence from
experiments in favour of CR, other physical effects of the gravity of the
Universe, especially on quantum systems, and proposal for experimental tests
with observable differences from predictions of SR. Part of the paper contains
results that can be considered as the quantum generalization of the Mach's
principle, but calculated directly from cosmology and a new theory of relativity.

\section{Gravity of the Universe and its effect on moving clocks}

\subsection{Physical and mathematical framework}

I now present the physical and mathematical reasoning behind the calculation
of the gravitational effects on clocks moving in the cosmos. This is discussed
before the full theory is discussed to contrast the new theory with SR on the
basis of already available experimental results on rates of moving clocks.

In a homogenous and isotropic Universe with an `age' of about 14 billion
years, the gravitational potential (energy) at any point is finite when
calculated as the integral effect of the potential from all the masses upto
the Hubble radius. There is no spatial gradient of this potential and there
are no forces. Our fundamental hypothesis of the theory, argued later, is that
this average gravitational potential determines a maximum velocity for any
entity moving in this cosmos, and that  this constant velocity is numerically
related to the potential as $c^{2}=\left|  \phi_{U}\right|  .$ In a
relativistic theory with four-potentials or more complex structure, moving
though such a cosmos will generate both a different gravitational potential,
$\phi(1-V^{2}/c^{2})^{-1/2},$ and also a vector potential, $-\phi\frac{V}%
{c}(1-V^{2}/c^{2})^{-1/2}$. Here $V$ is the velocity relative to the average
rest frame defined by all the masses in the Universe. Clearly, there is one
preferred frame in which the magnitude of the $\phi$ potential is minimum. In
all other frames moving relative to such a cosmic rest frame, the
gravitational potential has higher magnitude, and therefore all moving clocks
will be gravitationally slowed down compared to a clock that is at rest with
respect to the cosmic rest frame. Similarly, the vector potential will
interact with mass currents, and will cause interesting effects like spin
precession and phase changes on quantum systems. We will come to those effects
later in the paper.

Mathematically, one can start from the Robertson-Walker line element of the
homogenous and isotropic Universe.
\begin{equation}
ds^{2}=-c^{2}dt^{2}+\frac{R^{2}(t)}{1-kr^{2}}(dr^{2}+r^{2}d\Omega^{2})
\end{equation}

Since it is known from observations that the present Universe is nearly flat
in spatial sections, $k\simeq0.$ Then the line element is
\begin{equation}
ds^{2}=-c^{2}dt^{2}+R^{2}(t)(dr^{2}+r^{2}d\Omega^{2})
\end{equation}
As mentioned earlier $R(t)$ is negligible for time scales of years, and can be
approximated as a constant even for extended laboratory experiments. Then we
have
\begin{equation}
ds^{2}\simeq-c^{2}dt^{2}+R^{2}(dr^{2}+r^{2}d\Omega^{2})
\end{equation}
This is same as Minkowski flat space-time in the special frame (cosmic rest
frame) that is at rest with respect to the average Hubble flow. The time
coordinate $t$ is an absolute time applicable to all observers at rest with
respect to that special cosmic frame, and the CMBR temperature is a monotonic
function of $t$.

It is easy to see that a clock moving relative to the average matter
distribution will have a slower rate. We consider a clock in a frame moving
with physical velocity $V=R\frac{dr}{dt}$ with respect to the cosmic rest
frame. In a frame moving with the clock,
\begin{equation}
ds^{\prime2}=-c^{2}dt^{\prime2}%
\end{equation}
The same clock, with reference to the global frame has the interval,
\begin{equation}
ds^{2}=-c^{2}dt^{2}+R^{2}dr^{2}%
\end{equation}
Considering radial motion for simplicity (any linear motion can be treated
this way) and equating the intervals (see later for a proof that the interval
is invariant),
\begin{equation}
dt^{^{\prime}2}=dt(1-V^{2}/c^{2})^{1/2}%
\end{equation}
Since $\int dt$ is the time measured in the global frame, the moving clock
advances slower by the factor $(1-V^{2}/c^{2})^{1/2}.$

Thus a clock moving with respect to the cosmic rest frame records less time
compared to a clock that is stationary in the cosmic rest frame. The time
dilation is a gravitational effect, and always involves the velocity with
respect to a preferred cosmic rest frame. Note that there is no assumption
that the motion is uniform and inertial. The expression is applicable to
arbitrary motion, with $V=V(t).$

Our only assumption is that the velocity of light is an absolute fundamental
constant in the cosmic rest frame, determined physically by the gravitational
interaction of light with the Universe. We will show later that this
\emph{implies} the Lorentz transformations and that the velocity of light is a
constant in all frames.

\emph{It is very important to realize that the effects are gravitational in
origin and that in an empty Universe there will not be any relativistic
effects, unlike in SR}. It is possible to get convinced of this by considering
the effect of the distant galaxies on local physics. One starts by adding up
the individual potentials and by calculating the approximate metric, including
all matter visible to observational astronomy. This of course does not
represent all matter and all of the Universe, but gives a metric that is
different from the special relativistic Minkowski metric with metric
coefficients that are different from unity by $2\Phi/c^{2},$ where $\Phi$ is
the total gravitational potential of all visible galaxies, and $2\Phi/c^{2}%
\ll1.$ The next step is to calculate the effects of moving through this
gravitational potential, by using any post-Newtonian framework, or some
approximation to General Relativity, for example. \emph{Then one realizes that
these are substantial effects not seen in experiments}. Considering only
visible matter constituting 5\% of the critical density in the Universe, a
moving clock has a gravitational time dilation that is 5\% of the value
predicted by special relativity, relative to a clock stationary with respect
to the CMBR. For example, one may calculate the gravitational time dilation
effect of the galaxies on a muon circulating in a storage ring for which the
measurement of modification of lifetime agrees with the SR prediction to
0.1\%. If the gravitational potential due to the galaxies in the cosmic rest
frame is designated as $\Phi_{g},$ we have $\Phi_{g}/c^{2}\simeq0.05.$ In a
frame moving at velocity $V$ relative to the cosmic rest frame (CMBR), the
potential is
\begin{equation}
\Phi_{g}^{\prime}(V)=\Phi_{g}(1-V^{2}/c^{2})^{-1/2}\simeq\Phi_{g}%
(1+V^{2}/2c^{2})
\end{equation}
Therefore the gravitational time dilation of the moving clock relative to the
stationary clock will be
\begin{equation}
\Delta T\simeq-\frac{\Phi_{g}}{c^{2}}\frac{V^{2}}{2c^{2}}%
\end{equation}
With $\Phi_{g}/c^{2}\simeq0.05,$ this is 5\% of the special relativistic time
dilation, $-V^{2}/2c^{2}.$\ Such a large effect is not seen in any experiment,
sensitive typically to 0.1\% corrections. Clearly, \emph{there is no room for
both the gravitational effect and the special relativistic effect}. Thus, the
muon storage ring experiment is already in conflict with SR prediction based
on mere kinematics in empty space. One also sees that if one extrapolates the
results to include gravitational effects of all matter, integrating upto the
Hubble radius, then the physical effects are as large as those seen in
experiments, and attributed usually to special relativistic kinematical
effects. Since one does not want to conclude that gravitational effects from a
visible Universe predict something that is not seen in experiments, the only
way out is to admit that all the effects that are attributed to kinematics and
SR are indeed the gravitational effects of the entire Universe, and that SR is
an approximate effective description that is applicable in certain special
cases of motion (see later). The more fundamental theory -- Cosmic Relativity
-- is based on the gravitational effects of the Universe and it is not limited
to reference frames moving with uniform velocity.

\subsection{Results for clock comparison experiments}

Now I will discuss the time dilation effects within cosmic relativity.

Consider a frame moving at velocity $V$ with respect to the cosmic frame. We
consider experiments in which there are clocks moving within this frame, which
will be compared among themselves and with other clocks that are at rest
within the frame.

Consider a clock within this frame that is moving at velocity $u$ relative to
the coordinates inside the frame. The special relativistic time dilation of
the clock with respect to a clock that is stationary inside the frame is
\begin{equation}
dt^{\prime}=dt(1-\frac{u^{2}}{c^{2}})^{1/2}%
\end{equation}
In fact all clocks that move with a speed $u$ relative to the clocks
stationary inside the frame will record the same time dilation factor
irrespective of the direction of $u,$ according to special relativity. If the
clocks are accelerating, the total time dilation is given by
\begin{equation}
\int dt^{\prime}=\int dt(1-\frac{u(t)^{2}}{c^{2}})^{1/2}%
\end{equation}
In particular if two clocks are moving with velocities $\overrightarrow{u(t)}$
and $-\overrightarrow{u(t)}$ relative to the frame, the time dilation factors
of both clocks are identical in Special Relativity. There is no difficulty in
dealing with time dilation of noninertial clocks, since the total effect can
be calculated by integrating differential effects in instantaneous Lorentz
frames that are inertial. SR asserts that no special relativistic time
dilation expression should contain the velocity of the frame (with respect to
some hypothetical frame in which the moving frame is embedded) in which the
experiment is performed.

Just to stress this point I quote from the creator of the theory and the
editor who encouraged it.

\noindent\noindent A. Einstein (1905, \cite{ein-sr}): ``If one of two
synchronous clocks at A is moved in a \ closed curve with constant velocity
until it returns to A, \ the journey lasting $t$ seconds, then by the clock
that has remained at rest the travelled clock on its arrival at A will be
$\frac{1}{2}tv^{2}/c^{2}$ second slow''. There is no reference to closed curve
being noninertial, direction of velocity etc. The velocity in SR refers to the
relative velocity.

\noindent\noindent M. Planck (1909, \cite{planck-lect}): ``The gist of the
principle of relativity is the following. It is in no wise possible to detect
the motion of a body relative to empty space; in fact, there is absolutely no
physical sense in speaking about such motion. If, therefore, two observers
move with uniform but different velocities, then each of the two with the same
right may assert that with respect to empty space he is at rest, and there are
no physical methods of measurement enabling us to decide in favour of one or
the other''. \ 

When these were written there was not much knowledge on cosmology or on the
average density of the Universe. Since space it not empty, though it is flat,
effects that go beyond special relativistic assertions could naturally be expected.

The cosmic gravitational time dilation has the characteristic imprint of the
fact that there is a preferred cosmic frame with reference to which the time
dilation is calculated. The clock that is stationary within the frame itself
has a time dilation with respect to the clocks in the cosmic rest frame. (Such
a clock is provided by the temperature of the CMBR). We have to calculate the
time dilation of the stationary clock and the moving clock with respect to the
cosmic frame (subscript $U$) and then compare them. For the clock stationary
inside the frame (denoted by subscript $0$) this is
\begin{equation}
dt_{0}=dt_{U}(1-\frac{V^{2}}{c^{2}})^{1/2}%
\end{equation}
and for the moving clock the time dilation (after approximating
$(V+u)/(1+Vu/c^{2})$ as $V+u$ for $V,u\ll c$, see later) is%
\begin{equation}
dt_{1}=dt_{U}(1-\frac{(\overrightarrow{V}+\overrightarrow{u})^{2}}{c^{2}%
})^{1/2}%
\end{equation}
The time dilation of the moving clock with respect to the reference clock
inside the frame then is%
\begin{equation}
dt_{1}=dt_{0}(1-\frac{(\overrightarrow{V}+\overrightarrow{u})^{2}}{c^{2}%
})^{1/2}/(1-\frac{V^{2}}{c^{2}})^{1/2}%
\end{equation}
When both $V$ and $u$ are small compared to $c$, we get%
\begin{align}
dt_{1} &  =dt_{0}(1-\frac{V^{2}+u^{2}+2\overrightarrow{V}\cdot\overrightarrow
{u}}{c^{2}})^{1/2}/(1-\frac{V^{2}}{c^{2}})^{1/2}\nonumber\\
&  \simeq dt_{0}(1-\frac{u^{2}+2\overrightarrow{V}\cdot\overrightarrow{u}%
}{2c^{2}})
\end{align}
This is drastically different from the special relativistic time dilation by a
factor $dt_{0}(\overrightarrow{V}\cdot\overrightarrow{u}/c^{2})$. The
surprising new result is the dependence of the time dilation factor on the
velocity of the frame. This is equivalent to considering all velocities
relative to the cosmic rest frame or CMBR for calculating the time dilation
effect. In SR, no local experiment should have a dependence on the velocity of
the frame. Note that it is possible to have the factor $\overrightarrow
{V}\cdot\overrightarrow{u}/c^{2}$ negative and numerically larger than
$u^{2}/c^{2},$ and therefore it is possible to have a moving clock inside a
local frame \emph{age faster} than stationary clock in the same frame, in
complete contrast to the special relativistic prediction. Thus, it becomes an
issue of experiments to decide which of the two predictions is correct.

(It may be noted that the usual Lorentz transformations of time contain a
factor $Vx/c^{2},$ due to the finite velocity of light used in clock
synchronizations, which is structurally same as the factor we have derived for
time dilation. Therefore, this effect can never be observed in clock
comparisons done at a distance, since the effect depends on the direction of
travel and cancels out in two-way experiments. But, the effect reveals itself
in special experimental situations.)

The expressions above means that if two clocks are compared, one moving with
speed $u$ along the direction of $\overrightarrow{V},$ and the other with
speed $u$ in a direction opposite to $\overrightarrow{V}$, there will be a
\emph{time asymmetry} when the two clocks are compared \emph{with respect to
each other }by an asymmetry factor
\begin{equation}
\eta=\frac{t1-t2}{t_{av}}=\frac{2Vu}{c^{2}}%
\end{equation}
This multiplied by the duration of the experiment gives the nett time
asymmetry. The velocity $\overrightarrow{V}$ here is the velocity of the frame
with respect to the cosmic frame, and for a laboratory on earth, this will
include the rotational, orbital, solar system and galactic velocities. The
instantaneous velocity of any laboratory frame on earth's surface is time
dependent and it is the vector sum of the different contributions to its
velocity -- rotational ($R\Omega),$ orbital ($V_{orb}),$ solar system/galactic
velocity with respect to the cosmic back ground ($V_{U})$ etc. So the total
velocity is
\begin{equation}
\overrightarrow{V}(t)=R\Omega\widehat{\phi}+\overrightarrow{V}_{orb}%
+\overrightarrow{V}_{U}+...
\end{equation}

The time dilation asymmetry is proportional to $\overrightarrow{V}%
\cdot\overrightarrow{u}_{clock}.$ But, in any experiment where either
$\overrightarrow{u}$ takes both positive and negative values symmetrically, or
all directions with respect to the cosmic frame as in a storage ring for
unstable particles, the dot product averages to zero. Therefore, two way
experiments in which $\overrightarrow{u}$ and $-\overrightarrow{u}$
symmetrically occur with $\overrightarrow{V}(t)$ not changing direction, as
well as storage ring type of experiments on clocks will not see any time
dilation asymmetry.

If a clock is taken around the earth along the equator at constant ground
speed $u$ , and brought back after a round trip, \emph{its time dilation with
respect to a clock stationary on the surface is not given by the special
relativistic factor predicted by Einstein in 1905},%
\begin{equation}
\Delta T=-T\frac{u^{2}}{2c^{2}}%
\end{equation}
The correct result, according Cosmic Relativity should be
\begin{equation}
\Delta T=-T(\frac{u^{2}}{2c^{2}}\pm\frac{V_{rot}u}{c^{2}})
\end{equation}
The $+$sign applies when the clock is taken eastwards, along the surface
velocity of the earth due to its rotation, and $-$sign for the trip westwards.
All other asymmetry terms are zero. For example the term $\overrightarrow
{V}_{Orb}\cdot\overrightarrow{u}$ averages to zero since $\overrightarrow{u}$
takes all directions in a round trip along the equator.

Most interestingly, the second term can dominate with a negative sign, and
then we get the result that \emph{the clock in motion ages faster than the
clock at rest} within the frame. This can never happen in SR.

For a clock moving at ground speed $u$ along the instantaneous surface
velocity (440 m/s) of the rotating earth ($V=R\Omega$)with respect to the
cosmic frame, and another one moving opposite (eastwards and westwards) the
time asymmetry factor is
\begin{equation}
\eta=\frac{2Vu}{c^{2}}\simeq10^{-14}u
\end{equation}
where $u$ is in m/s. If the clocks are taken around in aircraft with a
velocity 220 m/s (average ground speed of about 800 km/hour), for about 40
hours for the round trip, then the predicted asymmetry would be
\begin{equation}
\Delta T=\eta T=\frac{2R\Omega uT}{c^{2}}\simeq310~\text{ns}%
\end{equation}
This is several times larger than the special relativistic time dilation,
$\Delta t\simeq-t_{av}u^{2}/2c^{2}\simeq50$ ns.

If an experiment is performed such that the clocks start moving eastwards and
westwards around noon (or midnight), then their velocities are parallel and
antiparallel to the orbital velocity of the earth around the sun. After a
short flight of 1 hour the Cosmic Relativity predicts a time asymmetry of a
whopping 720 ns, for a ground speed of 220 m/s. But for a neat comparison we
need to bring the clocks back. This will cancel the effect, and therefore the
comparison has to be done at a distance. This is not possible since
synchronizing clocks at a distance will contain the same terms we are trying
to detect, and therefore the only way such experiments can be done is by
bringing the clocks back in a round trip around the earth in a constant speed orbit.

Though not obvious from the way we started our analysis, a look at the
expression for $\Delta T$ indicate that the total time dilation asymmetry
depends only on the product $uT$, which is just the total path length covered
in the experiment. It does not matter how fast the clocks are moved, provided
we move them by the same distance. Slow transport will need more time, and the
asymmetry depends only the product of the velocity and duration. Thus if the
clocks are taken around by \emph{walking around the earth} eastwards and
westwards along \ equator, the clocks will show an asymmetry that is exactly
equal to the one predicted for clocks taken around in fast flights! The only
requirement for a good measurement is that the clocks be stable over a long
time. In contrast the special relativistic time dilation that is quadratic in
the velocity will be negligible for the slowly transported clocks. For
example, at a transport velocity of 1 m/s, around the earth, the special
relativistic time dilation is \ only a fraction of a nanosecond, whereas the
time dilation asymmetry between the two clocks is still about 310 ns.

There is strong and unambiguous evidence for the gravitational effect of the
Universe, or for the validity of Cosmic Relativity, from earlier clock
comparison experiments.

\section{Experimental evidence for Cosmic Relativity}

As I have explained in earlier sections, we need to look at an experiment in
which the term $\overrightarrow{V}\cdot\overrightarrow{u}_{clock}$ does not
average to zero to test whether the correct theory is SR or CR. One of the
classic tests of SR with clocks is the modification of the lifetime of
unstable particles like muons when they move at large velocity. In cosmic ray
experiments, the accuracy of lifetime measurement is not better than a percent
or so, and usually there is averaging over all directions. Thus the asymmetry
term averages to zero. A more precise experiment (0.1\%) is the lifetime
measurement of muons in a storage ring. But, again the asymmetry term averages
to zero, since the muon velocity takes all directions in the storage ring with
respect to the velocity of the earth in the cosmic frame. I have already
mentioned how this measurement indicates a conflict between gravitational
physics and SR.

It turns out that (apart from GPS whose precision data is not publicly
available) there are only two experiments with enough accuracy that can throw
light on the question we are discussing. I will discuss the famous clock
comparison experiment by Hafele and Keating in 1971-72 \cite{h-k}. Later clock
comparison experiments with higher precision \cite{alley} confirm these observations.%

\begin{figure}
[ptb]
\begin{center}
\includegraphics[
height=2.5547in,
width=3.2465in
]%
{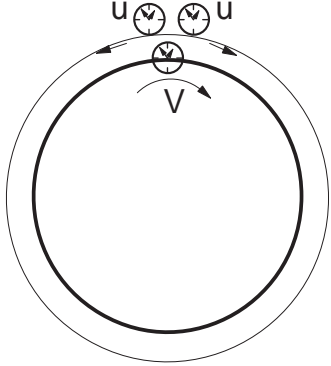}%
\caption{Round trip clock comparison}%
\end{center}
\end{figure}

Hafele and Keating flew four clocks in commercial aircraft around the earth,
one set eastwards and another westwards. When they reached back, they were
\emph{compared with a clock stationary on earth's surface}. The transported
clocks read times different from that of the stationary clock as expected, but
in addition the transported clocks also showed a large time asymmetry between
them. The average total duration of the round trip was about 45 hours. The
clock that was transported westwards was found to be advanced (`older') with
respect to the stationary clock, and the others that flew eastwards were found
to be retarded (`younger'). When compared with the stationary clock, one has
to take into account of the gravitational redshift due to the fact that the
aircraft clocks are at a different distance from the earth's centre compared
to the one on the surface and there is a large gravitational time dilation.
The flight clocks will run faster than the surface clocks by about 150 ns for
the parameters of the experiment due to the gravitational effect of the earth.
Special relativistic time dilation predicts that the flight clocks will run
slower by about 50 ns after the flight. Both these effects are symmetrical for
the two clocks in flight. Yet, Hafele and Keating found that one clock gained
time whereas the other lost compared to the stationary clock.

The details of the flights were as follows:

The eastward flight lasted 41 hours, and the calculated gravitational effect
in earth's field is faster aging by 144 ns relative to the surface clock. The
westward flight lasted 49 hours, and the gravitational effect is 179 ns.
Special relativistic effect calculated using the prescription in Einstein's
original paper, $\Delta t\simeq-tv^{2}/2c^{2},$ is $-36$ ns for the eastward
clock and $-45$ ns for the westward clock, the 20\% difference coming from the
20\% difference in flight times. So, the expected time difference relative to
the stationary clock at earth's surface is 108 ns for the eastward clock and
134 ns for the westward clock, with about 20\% errors in estimation. The
observed time differences were $-59$ ns eastwards, and $+273$ ns westwards.
After subtracting out the gravitational effects in the earth's field, it is
found that the clock that travelled eastwards aged less relative to the
surface clock by about 203 ns, and the clock that travelled westwards
\emph{aged more relative to the stationary clock} by as much as 94 ns.
\emph{Both these numbers are very different from the prediction of aging less
by about 40 ns, calculated as in Einstein's 1905 paper}. The `travelled clock
aging more than the stationary clock' contradicts the special relativistic
prediction. Significantly, these numbers agree within errors (20\% in
estimates of flight parameters and 10\% in experimental data) with what one
would expect if one clock travelled at a velocity of about 660 m/s for 41
hours and another at velocity of about 220 m/s for 49 hours, while the
reference clock itself travelled at a velocity of about 440 m/s (linear
velocity of the earth's surface during flight take-off), relative to the
cosmic rest frame.

\emph{It is important to note that the reference clock on the earth's surface
is irrelevant for the analysis of the time dilation asymmetry between the two
clocks.} For a direct comparison of the clocks in flights, the clocks on
earth's surface does not come into picture. The two clocks in flights execute
nearly symmetrical motion with respect to each other, and when they are
compared with each other, there should not be any asymmetry between them
according to SR. But an asymmetry of about 300 ns was found between the two clocks.

The authors unfortunately went through an inconsistent analysis of their data
to establish that Special Relativity was valid \cite{h-k,hafele}. They argued
that they should not compare their moving clocks with the clock stationary on
earth's surface because such a clock would not obey SR due to its noninertial
motion. They denied the applicability of instantaneous Lorentz transformations
on clocks moving with constant speed, with a continuous change in the
direction of velocity. They calculated the expected effect by comparing the
clocks in aircraft, and the stationary clock on earth with a hypothetical
clock that is moving with the center of mass motion of earth, but nonrotating
with respect to the cosmic frame. Equivalently, they went to a cosmic frame
(see later) and compared all clocks exactly as in Cosmic Relativity. Though
they did go to an external super-frame to use `inertial reference clocks', the
clocks in the aircraft, which executed motion very similar to the clock on the
earth was treated as clocks obeying SR -- a surprising and obvious
inconsistency. Thus they used the argument that clocks on earth's surface were
noninertial and not good as reference, \ but ignored that consistency demands
that the clocks that are to be compared also have to obey the same criterion
by SR. The fact of correct physics is that there is no need to go to the
external hypothetical frame in SR. It has been established in particle
accelerator experiments that the noninertial nature from circular motion is
not a problem in applying Lorentz transformation equations. Lorentz
transformation can be used in accelerating and rotating frames, by using
instantaneous Lorentz frames and then integrating the effect. Thus, an
asymmetry will be observed only if Special Relativity is not the correct
theory of time dilation.

Let us list the reasons why the Hafele-Keating analysis is inconsistent and
their claim of having verified special relativity is incorrect:

\begin{enumerate}
\item The two clocks in flight execute identical motions \emph{relative to
each other}. Therefore, they should not show any time asymmetry with respect
to each other in SR. Such an asymmetry is a prediction of Cosmic Relativity.
The observed \emph{asymmetry} is independent of whether it is obtained by
comparing with clocks on earth's surface, or with some hypothetical clocks
that are nonrotating, or \emph{without considering any other reference
clocks}, invalidating the entire reasoning given by Hafele and Keating for
establishing validity of SR.

\item In SR, there is clear prediction that if the clock is moved in whatever
way with respect to a stationary clock, whether in an inertial or noninertial
frame (again using instantaneous Lorentz frames) the clock that is moving will
\emph{always slow down with respect to the stationary clock}. This is
precisely the prediction in Einstein's original paper. There is no way in
special relativity that it will age more than a stationary clock. But
Hafele-Keating result shows that this happens. Such an effect is the
prediction of Cosmic Relativity.

\item The clocks can be compared by taking them around at very slow speeds,
and the time dilation asymmetry will remain the same. The special relativistic
time dilation, quadratically dependent on the velocity, can be made negligible
by moving slow.

\item The experiment actually compared the real noninertial clocks in flights
with the real noninertial clocks on earth, and not with inertial hypothetical
clocks. So, bringing in inertial hypothetical clocks as \emph{essential}
intermediaries for calculation points to the presence of the cosmic
gravitational effects and to the inadequacy of SR.

\item Suppose the earth was neither rotating nor moving in the solar system.
The two flight going eastwards and westwards at the same ground speeds have
identical and symmetrical circular paths with respect to each other. Therefore
there cannot be any time asymmetry between them. The relative aging of the
two clocks when they are compared should be an invariant physical fact. If we
observe this situation from an external rotating frame, rotating at the
angular velocity of the earth, then we have the Hafele-Keating situation.
According to SR such a common rotational velocity should not change any result
on clock comparisons. But there is a real physical difference in the Machian
sense, between the earth and the flights rotating together with respect to the
preferred rest frame and gravitational configuration of the cosmos, and a
situation where these are not rotating, but the observer is rotating. There
will be a corresponding difference in physical effects in Cosmic Relativity.

\item Finally we note that no relativist analyzes the standard twin-clock
problem in SR by going to the preferred cosmic frame encompassing the two
twins' motion, though it is clear that choosing such a frame in which the
stationary twin is `absolutely' stationary and the travelling twin is in
motion relative to the cosmos immediately gives the result that the travelling
twin ages less than the stationary twin. Nobody does it this way because
choosing such a frame denies and discredits SR, and implies that the motion
should be analyzed from a frame in which the concepts `stationary' and
`moving' has absolute sense. But Hafele and Keating did exactly this. Once you
go out of the local frame to a super frame (the nonrotating absolute frame),
one of their clocks move faster than the other, in an absolute sense, during
the round trip. But this is Cosmic Relativity and not Special Relativity.
\end{enumerate}

Hafele and Keating observed a time dilation asymmetry of 332 ns with a
statistical uncertainty of less than 4\%. The observed asymmetry matches well
with the time asymmetry predicted by Cosmic Relativity.

\emph{Thus there is no doubt that the result observed by Hafele and Keating is
due to the gravitational effect of the Universe on moving clocks and not due
to kinematics of relative motion as described by special relativity
}\cite{unni-hk}.

Similar evidence will be present in GPS data. The correct synchronization of
these clocks to good precision will have to use results from Cosmic Relativity
presented here. What is usually called the Sagnac term in this context is the
gravitational effect of the Universe. This will be true also for the planned
European satellite and clock network. 

\section{New tests of Cosmic Relativity}

Though we expect modification of all physical measures like time, length, mass
etc., due to gravitational effects of the Universe, the best experiments for a
precision test are clock comparison experiments.

A new experiment that repeats the Hafele-Keating experiment with an
improvement of precision from 10\% to 1\% is certainly very important. But,
that is a tedious experiment and the effect is not large since the only frame
velocity that contributes to the time dilation asymmetry is the rotational
velocity of the earth's surface, 440 m/s. But, as mentioned earlier, the
gravitational effects we are discussing are universal and any scheme of
comparing or synchronizing clocks at a distance will have the asymmetry term
relevant for the separation of the clocks $(Vx/c^{2})$, and the only clean way
of doing an experiment is to depend on the round trip comparison. Therefore we
clarify certain issues in round trip comparison experiments to unambiguously
show that the only consistent interpretation of the Hafele-Keating type
experiment and also the original twin clock problem is possible only within
Cosmic Relativity and not in Special Relativity.

\subsection{Analysis of a variation of the Hafele-Keating experiment and proof
for the validity of Cosmic Relativity}

Consider the race-track clock comparison experiment depicted in figure 2.%

\begin{figure}
[ptb]
\begin{center}
\includegraphics[
height=1.6198in,
width=4.0387in
]%
{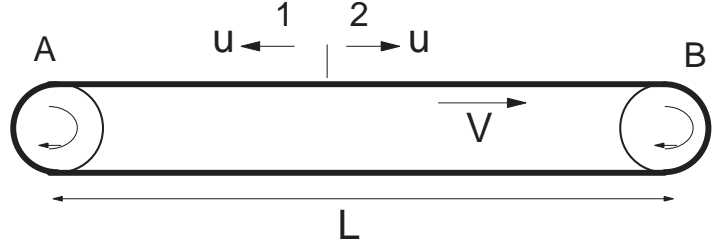}%
\caption{Analyzing this experiment reveals the inconsistency and inadequacy of
Special Relativity}%
\end{center}
\end{figure}

The experiment is done such that most of the motion of both the clocks are
straight inertial trajectories. The only portions where there are noninertial
motions are at the regions A and B. The clocks are transported at equal ground
speeds, $v_{g},$ relative to the track surface, and the track itself is like a
conveyor belt of total length $L,$ moving at velocity $v_{t}.$ The clocks
start from the same point, go around and come back and meet at the starting
point after the round trips. Since the clocks come back and compare the entire
flight data there is no consistent way to include fictitious gravitational
fields into the problem since the two clocks will never consistently agree on
such fictitious fields. The SR calculation for each clock now gives nearly
identical time dilations and the only differences, if at all, could be due to
physical effects at the regions where there is noninertial motion. But, the
contribution from these regions can be made arbitrarily small by making the
linear portion of the race-track long enough (or by keeping the clock readings
frozen during accelerated motion). The prediction from Cosmic Relativity for
the time asymmetry between the two clocks however is proportional to the sum
of the length of the round trip trajectories for the two clocks,
\begin{equation}
\Delta T=\frac{2\overrightarrow{V}\cdot\overrightarrow{u}T}{c^{2}%
}=\frac{2v_{t}L}{c^{2}}%
\end{equation}
Also, one of the `travelled clock' can age more relative to the `stationary
clock', when the velocity of the race-track belt relative to the cosmic rest
frame is more than the velocities of the clock relative to the track. There is
no way to get this result in SR without invoking absolute velocities with
respect to the preferred frame of the cosmos. More importantly, the time
asymmetry between the clocks cannot be attributed to any pseudo-potential or
noninertial features since the effect is dominated by motion in the inertial
regime. In the limit of the linear portion of the race track shrinking to
zero, we get the Hafele-Keating situation with $v_{t}=R\Omega,$ and then the
time asymmetry is
\begin{equation}
\Delta T=\frac{2R\Omega\times2\pi R}{c^{2}}=\frac{4\pi R^{2}\Omega}{c^{2}}%
\end{equation}
So, the asymmetry in the clocks is not physically related to noninertial
motion, since in the general case the effect is proportional to the path
length traversed in total during inertial and noninertial motion. The entire
effect is due to the gravitational interaction with the Universe as predicted
by CR. This result justifies well the validity of Cosmic Relativity.

\subsection{The Sagnac Effect}

The Sagnac effect was first discovered in optical interferometry. The phase
shift in a rotating planar interferometer with area $A,$ in which light
travels in two opposite paths and return to their starting point is given by%
\begin{equation}
\Delta\varphi=\frac{4A\Omega\omega}{c^{2}}%
\end{equation}
$\omega$ is the angular frequency of light, and $\Omega$ is the angular
velocity of rotation.

This expression is same as the expression for the time asymmetry in round trip
clock comparisons, since $\Delta\varphi=\omega\Delta T,$ and
\begin{equation}
\Delta T=\frac{4\pi R^{2}\Omega}{c^{2}}=\frac{4A\Omega}{c^{2}}%
\end{equation}
It is implied that the physical interaction responsible for the Sagnac effect
is the gravity of the Universe. We will discuss this in detail in another
paper \cite{unni-sagnac}. Here we merely note that the total equivalence of
the expression for the Sagnac effect for light and matter waves arises from
the fact that gravitational interaction is universal, and therefore the
\emph{Sagnac effect does not depend on the group velocity of waves} used in
Sagnac interferometry (this result is not intuitively obvious, for example in
a Sagnac interferometer that uses optical fibers, since the light pulse takes
more time to circle around and yet the time difference between the clockwise
and counterclockwise pulses is still given by the same equation.) This result
can de derived simply in CR as the phase change on waves due to the vector
potential generated due to motion with respect to the universe. The expression
is same as in electrodynamics.%
\begin{equation}
\Delta\varphi=\frac{E}{\hbar c^{2}}\oint\overrightarrow{A_{g}}\cdot
\overrightarrow{dl}%
\end{equation}
where $E$ is the energy of the interfering entity ($\hbar\omega$ for light)
\cite{unni-sagnac}.

\subsection{The inconsistency in SR}

I have already shown that there is evidence against SR from the Hafele-Keating
experiment, and even from other `relativistic' experiments done in the ever
present gravity of the Universe. While  SR is a correct effective description
of phenomena in a limited domain, there are situations accessible to
experiments where SR has no consistent answer. Another example that is very
instructive is that of the original twin clock problem or the twin paradox.
While many physicists will be dismissive about any new discussion on this
issue, anybody who has attempted to calculate the proper time in the frame of
the travelling twin will realize immediately that there is a problem and that
there is no consistent answer within SR.

In the frame of the stationary twin, the answer from SR and Cosmic Relativity
are identical.

The proper time for the stationary twin is given by
\begin{equation}
T_{1}=\int g_{00}^{1/2}dt=\int dt
\end{equation}
since the corresponding clock is at rest. In the frame of the second twin, his
proper time in SR is given by two parts, one corresponding to the part of the
journey that is inertial, and another part where the metric is different due
to noninertial accelerations generated for turning back.
\begin{equation}
T_{2}=\oint g_{00}^{1/2}(t)dt=2\int_{I}dt+2\int_{NI}g_{00}^{1/2}dt
\end{equation}
There is no way to estimate the second term in SR since local measurements
within the frame give only the value of the acceleration. So, even if it is
equated to a pseudo-gravitational field, the gravitational potential and the
local metric are not determined, unless one takes the rest frame of the cosmos
as the reference. In any case, whatever the contribution from that term is, it
can be made \emph{arbitrarily small compared to the first term} by making the
inertial leg of the journey longer and longer. So, the proper time calculated
in both frames becomes approximately equal within SR. Or, rather trivially,
the clock can be made to work only during the linear inertial leg of the
journey, `frozen' during accelerations, and finally compared after returning,
and then SR has no consistent and correct answer. Note that the experiment can
be designed such that the clock is made to work only during the outward
inertial journey, and frozen in reading and brought back, and still have a
time dilation relative to the stationary clock, always correctly predicted in
Cosmic Relativity, but never consistently given in SR. In such an experiment,
SR allows either clock to be treated as at rest during the inertial first part
of the journey, and has no unambiguous prediction for the time dilation. But
Cosmic Relativity predicts that the clock that travelled relative to the
cosmos ages less compared to the one that did not, without any ambiguity.%

\begin{figure}
[ptb]
\begin{center}
\includegraphics[
height=1.9009in,
width=4.9934in
]%
{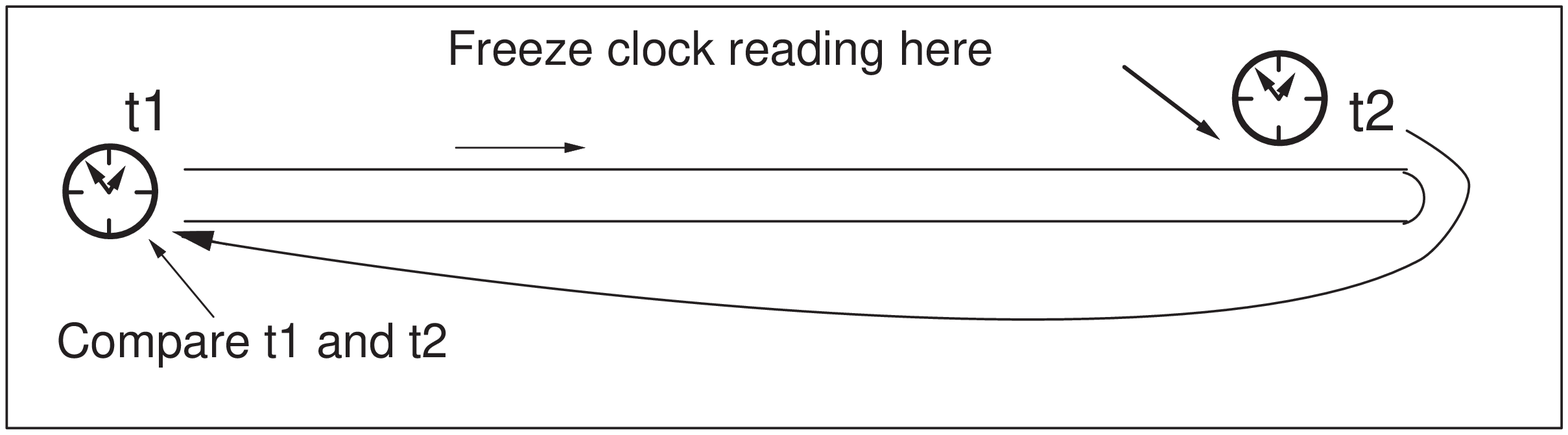}%
\caption{A variation of the twin clock problem. }%
\end{center}
\end{figure}

In Cosmic Relativity, the metric for twin 1 is $g_{00}=1,$ and for the twin 2
it is $g_{00}=(1-V^{2}/c^{2}),$ where $V$ is the velocity with respect to the
cosmic frame as determined by the dipole anisotropy in the CMBR in the frame
of twin 2. The effect during turn around is unimportant when the duration of
the uniform motion is long enough. Thus, the travelling twin ages slower due
to the gravitational effect of the Universe during inertial motion through the cosmos.

This analysis suggests that Cosmic Relativity is the only consistent theory of
relativity in flat space and time. SR that is based on kinematical effects in
relative motion fails in situations where comparison can be made after round
trips, with a major fraction of the round trip executed as inertial motion. In
such situations pseudo-potentials do not come to rescue since the non-inertial
contributions can be made arbitrarily small. Noninertial motion is treated the
same way as inertial motion in Cosmic Relativity.

Thus Cosmic Relativity is the generalized theory of relativity in flat
space-time, since it does not distinguish between inertial and noninertial motion.

\section{Cosmic Relativity and physical effects in quantum systems }

There are several important physical effects predicted by Cosmic Relativity,
all of which are manifestations of the gravitational interaction of the
physical system with the Universe. Two important effects we consider are the
spin precession in atoms and the phase changes in quantum systems that are
moving in the cosmos. Our results can be considered as the generalization of
the Machian idea to quantum systems. However, instead of starting from the
Mach's principle, we start from cosmology and relativity. For a critical
Universe, it turns out that the result on spin precession for a full orbit is
identical to the Thomas precession. We have identified the physical
interaction responsible for the Thomas precession -- it is the gravitational
interaction with the Universe \cite{unni-erice}. Thus part of the fine
structure splitting in atoms is due to the ever present gravitational
interaction of the orbital electrons with the massive Universe.

Consideration of phase changes of quantum systems yields a surprising and
pleasing result. The connection between spin and statistics in quantum theory
could be a consequence of the gravitational interaction of the spin with the
Universe. The interaction is gravitomagnetic in nature, and gives us the
result that identical integer spin particles obey Bose-Einstein statistics and
identical half-integer spin particles obey Fermi-Dirac statistics
\cite{unni-spin}. This is a deep result, and for the first time might answer
the long-standing query -- what is the physical reasons behind the
spin-statistics connection? It also answers why the connection is valid in
non-relativistic, two-particle situations despite the general impression that
it is a consequence of relativistic field theory.

\subsection{The spin precession in atoms and the fine structure}

It is perhaps not surprising that Cosmic Relativity will give the correct
spin-precession for electrons in atoms when we consider motion in a critical
Universe. As in the case of the muons in storage ring, the numerical results
from Cosmic Relativity matches the known results from SR in this case, for a
critical Universe. The available high precision data leave no room for both
kinematical effects and gravitational effects to be present together.
Therefore, since massive galaxies exist, we have to conclude that the observed
effects are indeed due to the gravitational effects of the Universe.

When the idea of electron spin was first proposed by Uhlenbeck and Goudsmit,
they had not resolved the problem that the simple L-S coupling gives twice the
experimentally observed value for the fine structure splitting. It is only
after Thomas derived the special relativistic Thomas precession \cite{thomas}
that things were set right (Though Kronig seems to have worked out the correct
relativistic expression earlier). The basic idea behind the Thomas precession
is that two Lorentz boosts in different directions are equivalent to a boost
and a rotation, and that this rotation is responsible for the precession of
the spin.

In Cosmic Relativity, the electron that is moving in the cosmic gravitational
potential $\phi$ of the Universe experiences a modified gravitational
potential and a vector gravitational potential equal to
\begin{align}
\phi^{\prime} &  =\phi(1-V^{2}/c^{2})^{-1/2}\\
A_{i} &  \simeq\phi\frac{V_{i}}{c}(1-V^{2}/c^{2})^{-1/2}%
\end{align}
Circular motion then gives a nonzero curl for the velocity field, and this is
a gravitomagnetic field due to the entire Universe. This gravitomagnetic field
couples to the spin angular momentum. Also, the modified potential modifies
precession rate of spin and this is exactly the Thomas precession
\cite{unni-erice}.

Thus we are able to identify the physical torque that is responsible for the
\emph{physical} precession of the spin, instead of attributing it to a
kinematical effect. It is certainly more satisfactory to identify a physical
cause for a physical change than to stop at a description of the physical
change. Since the Thomas precession term is usually written as
\begin{equation}
\omega_{T}=\frac{\mathbf{v}\times\mathbf{a}}{2c^{2}}%
\end{equation}
where $\mathbf{v}$ is the velocity and $\mathbf{a}$ the \emph{centrifugal
acceleration} in the orbit, it may be guessed that the term has something to
do with the gravitational interaction with the Universe. The fictitious forces
normally attributed to kinematics are in fact due to the gravitational effect
of the Universe \cite{sciama}. This effect is identical for a classical
gyroscope and a quantum spin, when we consider expectation values. This is
because the gravitational effects are universal, and just as the gravitational
acceleration is independent of the mass, the spin precession is independent of
the value of the spin itself. Experiments like Gravity Probe-B seek to measure
spin precession effect arising from a similar term in the gravitational field
$\mathbf{g}$ of the earth,
\begin{equation}
\omega_{g}=\frac{\mathbf{v}\times\mathbf{g}}{2c^{2}}%
\end{equation}
where, $\mathbf{v}$ is the velocity of the gyroscope through the Newtonian
gravitational field. Therefore, the only connections we need to realize that
the Thomas precession is due to the gravity of the Universe is the Machian
assertion that centrifugal forces in all situations arise due to the
interaction with the massive Universe, and the equivalence principle that
guarantees local equivalence of $\mathbf{g}$ and $\mathbf{a}$ \cite{unni-mpla}.

More explicitly, since the Thomas precession frequency for the nearly circular
orbit can be written as
\begin{equation}
\omega_{T}=\frac{\mathbf{v}\times\mathbf{a}}{2c^{2}}=-\frac{\mathbf{\Omega
}v^{2}}{2c^{2}}%
\end{equation}
it is clear that the term arises in the gravitational interaction of the spin
with the curl of the vector potential generated in moving through the critical
Universe,%
\begin{equation}
\nabla\times\mathbf{A}\simeq\Omega(1-v^{2}/c^{2})^{-1/2}%
\end{equation}

In the quantum treatment of this problem, the gravitational contribution to
the fine structure splitting is due to the quantized, two-valued coupling
energy between the quantum spin and the gravitomagnetic field of the Universe
generated by the orbital motion of the electron in the gravitational field of
the Universe. Thus, there is no need to invoke a classical torque acting on a
classical model of a spin.

Since the galaxies and massive Universe exist, the Dirac equation for the
electron should contain an additional term in the Hamiltonian that represents
the gravitational interaction of the electron with the Universe. As soon as
this term is included, gravitational effects of the type we discussed in this
section will be seen to emerge, including the correct Thomas precession term.
As I stressed earlier, precision experimental data do not allow both a
kinematical contribution from SR in empty space as well as the gravitational
effects from the Universe to be present together. The only possible physical
choice is to conclude that all measured relativistic effects are due to the
gravitational interaction with the Universe.

Thus, fine structure spectroscopy is the most precise measurement of whether
the Universe is evolving at the critical density \cite{unni-erice}.

\subsection{Gravity and the spin-statistics connection}

The spin-statistics connection in quantum mechanics is a very intriguing fact.
While there is no physical understanding of the connection, mathematical
proofs invoking field theoretic reasoning exist. The proof by Pauli used the
mathematical fact that the quantization of the field of integer spin particles
is associated with commutator relations and the quantization of half-integer
spin particles is associated with anti-commutator relations between operators,
and these coupled with requirements from Lorentz invariance provided the proof
\cite{pauli}. Later, there have been other attempts to provide simple proofs,
and the general assessment seems to be that there is no simple proof, let
alone a physical understanding of the connection.

First we state the spin-statistics connection:

a) Particles with integer spin are bosons and they obey the Bose-Einstein statistics.

b) Particles with half-integer spin are fermions and they obey the Fermi-Dirac statistics.

Equivalently (as normally attempted in proofs),

a) The amplitude for a scattering event between identical integer spin
particles and the amplitude with an exchange of the particles add with a plus
($+$) sign. In other words, the phase difference between the scattering
amplitude and the exchanged amplitude is an integer multiple of $2\pi.$

b) The amplitude for a scattering event between identical half-integer spin
particles and the amplitude with an exchange of the particles add with a minus
($-$) sign. In other words, the phase difference between the scattering
amplitude and the exchanged amplitude is an odd integer multiple of $\pi.$

A geometric understanding of these statements was published by Berry and
Robbins \cite{berry}, and several authors have invoked the relation between
rotation operators and exchange of particles in quantum mechanics to prove the
spin-statistic theorem \cite{spinstat-sud}. Sudarshan has been arguing for the
existence of a simple proof that is free of arguments specific to relativistic
quantum field theory\cite{sud}. While these attempts have clarified several
issues regarding the connection, none provides a good physical understanding
of the connection.

It may be noted that physically the connection is applicable for any two
identical particles, in non-relativistic quantum mechanics. Thus we should
expect that the physical proof need not depend on relativistic quantum field theory.

In what follows, we suggest that \emph{it is the gravitational interaction of
the quantum particles with the entire Universe that is responsible for the
spin-statistics connection }\cite{unni-spin}. In other words, the Pauli
exclusion is a consequence of the relativistic gravitational interaction with
the critical Universe, which is always present.

Consider the scattering of two identical particles, Fig. 4.%

\begin{figure}
[ptb]
\begin{center}
\includegraphics[
height=2.514in,
width=3.9211in
]%
{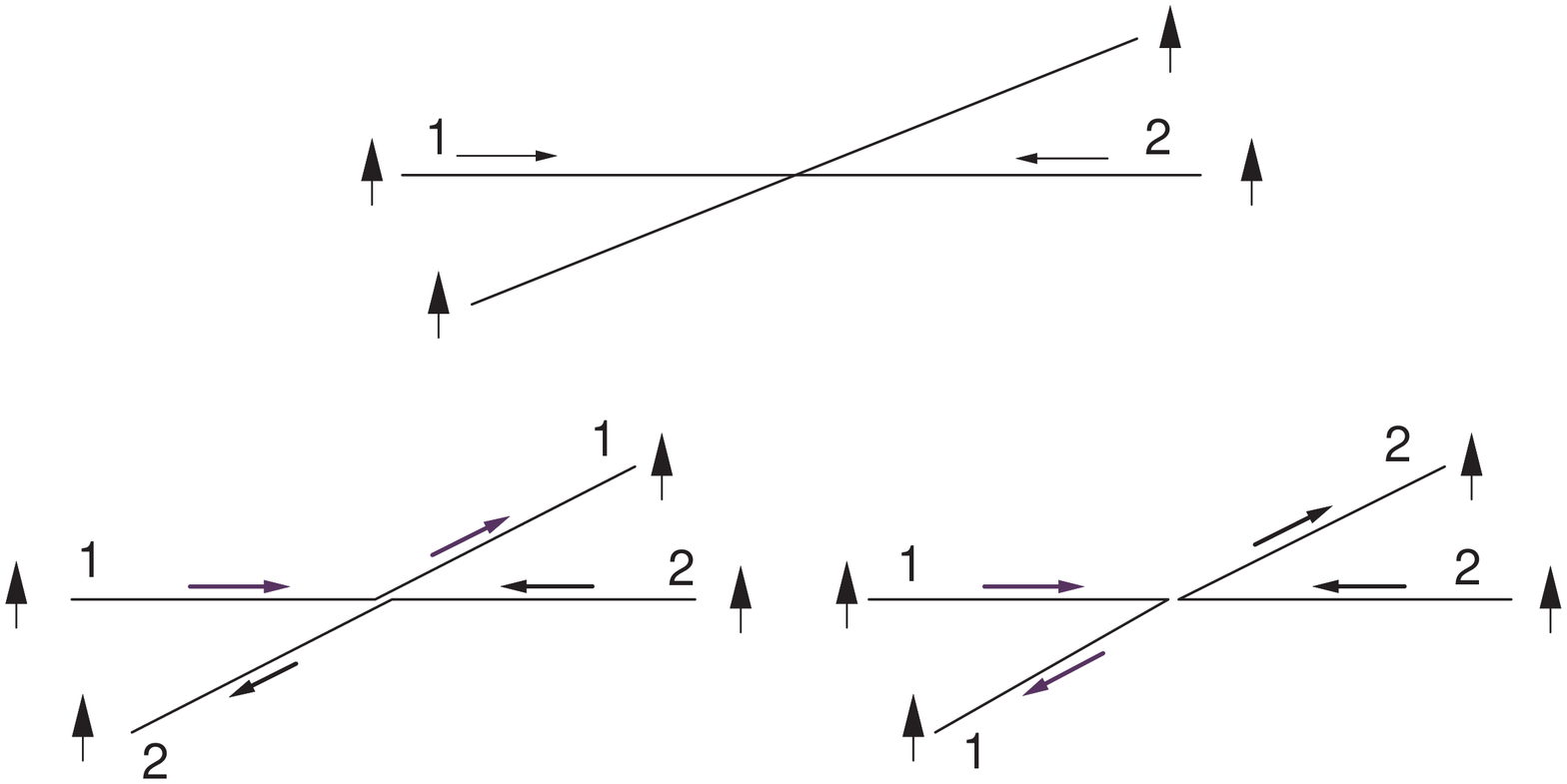}%
\caption{Amplitudes for scattering between indistinguishable particles}%
\end{center}
\end{figure}

The upper event can happen by two quantum amplitudes, shown in the lower
panel. The amplitudes for these two processes differ by only a phase for
identical particles in identical states, since the two processes are
indistinguishable. The particles are assumed to be spin polarized in identical
directions, perpendicular the plane containing the scattering event. Only
then, the initial and final states are indistinguishable. We can calculate the
phase changes in any configuration, but at present we want to discuss only the
phase difference for indistinguishable states.

First we note that the two processes are different in the angle through which
the \emph{momentum vector} of the particles turn in the scattering process. In
fact that is the only difference between the two amplitudes. The difference in
angles is just $\pi.$ (This is why it is equivalent to an exchange - what is
really exchanged is the momentum vector after the scattering). We note the
important fact that the entire process happens always in the gravitational
field of the entire Universe.

The calculation for each particle can be done by noting that the $k$-vector
can be considered without deflection, but the entire Universe turned through
an appropriate angle, with angular velocity of this turning decided by the
rate of turning of the $k$-vector. It does not matter whether we are dealing
with massless particles or massive particles since the only physical fact used
is the change in the direction of the $k$-vector of the particle.

As discussed earlier, the motion of the particle relative to the Universe
generates the vector potential, and the rotation of the momentum vector
generates a nonzero curl and therefore a gravitomagnetic field,
\begin{align}
A_{i}  & \simeq\phi\frac{V_{i}}{c}(1-V^{2}/c^{2})^{-1/2}\nonumber\\
\overrightarrow{B}_{g}  & =\nabla\times\mathbf{A}\simeq\overrightarrow{\Omega
}(1-v^{2}/c^{2})^{-1/2}%
\end{align}
where $\overrightarrow{\Omega}$ is the rate of rotation of the $k$-vector. If
we start this calculation from the Robertson-Walker metric, then what is
relevant is the curl of the vector potential which is just $\nabla
\times\overrightarrow{v}$ for a critical Universe. This field acts on the
particle only for the duration of the turning of the $k$-vector, and causes no
forces for a point particle. However, it can cause both spin-precession and
phase changes in a quantum particle. The spin-gravitomagnetic field
interaction energy is $\mathbf{s}\cdot\overrightarrow{\Omega}(1-v^{2}%
/c^{2})^{-1/2}.$ (This is what is responsible for the contribution to the fine
structure splitting in atoms). The phase change in the state of each particle
is given by the product of the duration of interaction and the
spin-gravitomagnetic interaction energy,
\begin{equation}
\varphi=\mathbf{s}\cdot\overrightarrow{\Omega}(1-v^{2}/c^{2})^{-1/2}\times
t(1-v^{2}/c^{2})^{1/2}=s\theta
\end{equation}
where $\theta$ is the angle through which the $k$-vector has turned in the
scattering process. The Lorentz factor correction to the vector potential does
not affect the final phase since it is the product of the gravitomagnetic
field and the time duration of interaction; the Lorentz factors cancel out.
Note that this is a gravitational phase shift, which involves all the
fundamental constants $G,$ $c$ and $\hbar.$ These constant are hidden from the
final expression because $\sum_{i}Gm_{i}/c^{2}r_{i}=1$ in a critical Universe.
When the density is not critical, these constants will appear explicitly in
the expression for this phase change. Also note that this phase difference is
an unambiguous prediction from relativistic gravity in flat space, in a
Universe with critical density. 

The momentum vectors turn in the same sense for both the particles and
therefore the total phase change is $\varphi_{1}=2s\theta_{1}$ where
$\theta_{1}$ is the angle through which the $k$-vector turns for the first
amplitude. For the second amplitude the phase change is
\begin{equation}
\varphi_{2}=2s\theta_{2}%
\end{equation}
The phase difference between the two amplitudes is
\begin{equation}
\Delta\varphi=2s(\theta_{1}-\theta_{2})=2s\times\pi
\end{equation}

The rest of the proof of the spin-statistics connection is easy. For zero-spin
particles the proof is trivial since
\begin{equation}
\Delta\varphi=s(\theta_{1}-\theta_{2})=0\times2\pi=0
\end{equation}
and therefore zero-spin particles are bosons and their scattering amplitudes
add with a $+$ sign. \ Zero-spin particles have no spin-coupling to the
gravitomagnetic field of the Universe and there is no phase difference between
the two possible amplitudes in scattering, and they behave as bosons.  What
remains is the proof for spin-half particles, since the higher spin cases can
be constructed from spin-half using the Schwinger construction. The only case
to be treated when dealing with identical, indistinguishable states of spin is
the one in which the spins are identically pointing, perpendicular to the
plane containing the $k$-vectors. The phase angle difference between the two
amplitudes then is%
\begin{equation}
\Delta\varphi=2s(\theta_{1}-\theta_{2})=2\times\frac{1}{2}\times\pi=\pi
\end{equation}
The relative phase is $\exp(i\pi)=-1.$ The amplitudes add with a negative
sign. Therefore, the half-integer spin particles obey the Pauli exclusion and
the Fermi-Dirac statistics. 

Our proof reproduces the result that exchanges of \emph{any} two particles in
a multiparticle system of identical fermions introduces a minus ($-$)  sign
between the original amplitude and the exchanged amplitude.

It is interesting to note that the proof is valid for interacting particles
since the phase changes due to interactions are identical for the two
amplitudes, as all other dynamical phases in the relevant diagrams.

This remarkable connection between quantum physics and gravity is indeed
startling. It is also very satisfactory physically since it is reasonable to
expect that a deep physical phenomenon is linked to a physical cause or
interaction and not just to mathematical structures and consistency. Most
importantly, as I always stress, the massive Universe exists, and the
gravitational phase difference between the two amplitudes is physically
unavoidable. If gravity is a long range interaction and if our understanding
of classical gravity is more or less correct, then these phase changes are
direct consequences.

Many of the measured geometric phases on particles with spin, when they are
taken around trajectories in space with their momentum vector turning in
space, are of the same nature. Those geometric phases are simply given by
\begin{equation}
\varphi=\oint sd\theta
\end{equation}
where I mean an integral over the entire trajectory with appropriate signs.
This clarifies to some extent the relation between our proof and the
discussion by Berry and Robbins. It also clarifies why the proofs that depend
on rotational properties of wavefunctions have some success. 

\subsection{Cosmic Relativity and Electrodynamics}

Static electromagnetic fields are modified by the gravitational potentials in
a laboratory moving relative to the cosmic rest frame. Detailed considerations
indicate a satisfactory description of phenomena like the unipolar induction
and its variations. There is no satisfactory description of such phenomena
within special relativity, especially since these seemingly counter intuitive
and asymmetrically relativistic phenomena occur in non-inertial situations. My
conclusion is that the gravity of the universe is the underlying interaction
that is responsible for the peculiar observations related to unipolar
induction. I will discuss these aspects in detail, along with some
experimental results and ideas, in separate papers
\cite{unni-erice04,unni-uni,gress04}.

\section{Principle of Cosmic Relativity}

Now I outline systematically the fundamental principle of Cosmic Relativity,
its relation to Lorentzian and Einsteinean relativity, and the intimate
connection to cosmology. While Lorentzian Relativity's transformation
equations are identical to the ones in Cosmic Relativity, the physical effects
are due to gravity in Cosmic Relativity and there is no ether. The entire
structure of Cosmic Relativity is based on flat space cosmology, as always
`given'. Thus, every physical effect that was tested as a special relativistic
effect was in fact the result from Cosmic Relativity, due to the gravitational
action of the entire Universe.

\subsection{The fundamental principle regarding the velocity of light}

The fundamental principle of Cosmic Relativity is that the velocity of light
is a fundamental constant determined by the local average gravitational
potential due to the entire Universe in the cosmic rest frame. (While I treat
this as an assumption of the theory, the restriction of velocities by the
gravitational potential can be derived once self-consistency is demanded.
Gravity of the Universe generates a phase drag on anything moving through the
Universe and limits the maximum possible velocity. See later). This is enough
to deduce that it is a constant in all frames. In SR, the constancy of
velocity of light in all frames is an assumption, whereas in Cosmic Relativity
it follows from the fact that it is constant in one special frame that
\emph{encompasses all frames within it}.

Since the cosmic rest frame encompasses everything moving within it, it is
clear that light emitted from a moving source will still have the same
velocity $c,$ irrespective of the state of motion of the source. Thus
identifying the preferred frame as the cosmic rest frame itself easily
accounts for the counterintuitive fact that the velocity of the source does
not affect the velocity of light emitted from it. This immediately implies
that the velocity of light measured in any moving frame also has to be an
invariant, $c.$ To see this, consider light emitted by any source outside the
frame (Fig. 5) entering a frame moving in the same direction with velocity
$\overrightarrow{v}$. The velocity of this light relative to the cosmic frame
is of course $\overrightarrow{c}.$ If the apparent velocity of light inside
the observer's frame moving at velocity $\overrightarrow{v}$ is
$\overrightarrow{c-v}$ then there will be a contradiction. The observer can
deflect the light in any direction inside his frame with a mirror stationary
in his frame, thus changing the direction of light without modifying its speed
inside his frame. But now, the speed of light relative to the cosmic frame
will not be $c.$ When the light is reflected back from a stationary mirror,
the apparent velocity inside the frame is $-\overrightarrow{(c-v)}.$ Then the
velocity with respect to the cosmic frame should be $-\overrightarrow
{(c-v)}+\overrightarrow{v}=-\overrightarrow{(c-2v)}$ instead of
$-\overrightarrow{c}.$ Therefore we can see that the existence of the absolute
cosmic rest frame, and the requirement that the velocity of light in that
frame is a fundamental constant imply that it is constant in all frames. In
Cosmic Relativity the numerical value is uniquely determined, and it is equal
to the average gravitational potential of the Universe at critical density.%

\begin{figure}
[ptb]
\begin{center}
\includegraphics[
height=2.0937in,
width=4.7236in
]%
{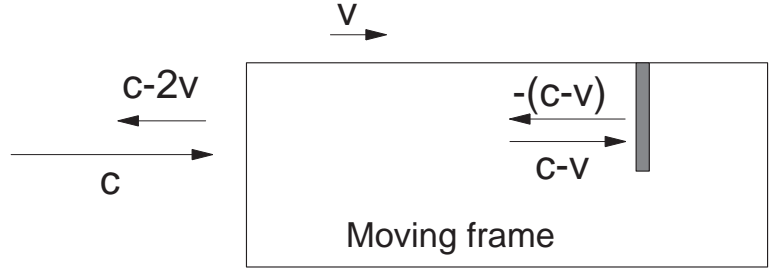}%
\caption{Relation between velocities in cosmic rest frame and moving frame if
the velocity of light were dependent on the velocity of the frame.}%
\end{center}
\end{figure}

From this it follows that the transformation equations connecting the cosmic
rest frame and any other frame moving within the Universe are the Lorentz
transformations. Lorentz transformations ensure that the velocity of light is
the same constant $c$ if measured in all frames. We do no have to assume the
constancy of velocity of light in all frames -- it follows from the fact that
the preferred frame encompasses all frames and observers. This is a remarkable
conceptual advantage of Cosmic Relativity over Special Relativity, apart from
its better agreement with cosmology and experimental facts. For the
consistency of the Lorentz transformations it follows that $c$ is also the
maximum velocity that can be reached from any velocity less than $c$ by
acceleration (I will address the issue of Tachyons elsewhere).

Since the transformation equations connecting the cosmic rest frame and any
other frame $S$ moving at velocity $V$ is the Lorentz transformation, the
standard velocity addition formula follows for the resultant velocity of a
second frame $S^{\prime}$ moving with velocity $u$ relative to the first
frame. The velocity of the frame $S^{\prime}$ with respect to the cosmic rest
frame when $V$ and $u$ are in the same direction is
\begin{equation}
U=\frac{V+u}{1+Vu/c^{2}}%
\end{equation}
Of course, this is approximately $V+u$ when $V,u\ll c,$ and we used this
approximation in the calculation of the clock comparison expressions.

\subsubsection{Universal gravity is the velocity limiting agency}

It is often asked what is the physical reason for a maximal velocity for any
entity in this Universe. The answer is available within Cosmic Relativity. It
is the gravity of the Universe that determines the numerical value of the
velocity of light, and it obeys the relation
\begin{equation}
\left|  \phi_{U}\right|  =c^{2}%
\end{equation}
In Cosmic Relativity, this is seen as a phase drag (not a force) due to the
gravitational potentials on any entity moving through the Universe.
Surprisingly, this view is consistent only if every material entity with a
gravitational interaction has also wave property and `phase' associated with
it. If the Universe were empty, the velocity of light would be infinite. The
actual numerical value is determined self-consistently; if we start by
assuming a large velocity, then the effective gravitational potential is
larger due to a larger causally connected Universe. More the gravity, more the
constraining effect on the phase advance of the light or any other entity.
Since the gravitational coupling is universal the restriction applies to light
as well as everything else.

\subsubsection{Why is E=mc$^{2}$ ?}

I have already mentioned the important physical relation between the velocity
of light and the average gravitational potential of the Universe at any point.
For the Universe at critical density,
\[
\left|  \phi_{U}\right|  =c^{2}%
\]

It is worth commenting on this relation. If the Universe started from pure
nothingness, then it is expected that every constituent of this Universe has
zero energy. One part of the energy is the gravitational interaction energy
given by $E_{g}=-m\phi_{U}.$ Clearly, every mass at rest with respect to the
cosmic frame should possess an energy, positive and equal to $E_{g}$ to be
consistent with its origin from void. Thus we expect that in a massive
Universe all masses possess a rest energy equal to $E_{0}=m\left|  \phi
_{U}\right|  .$ We write $E_{0}=\kappa mc^{2},$ since $\left|  \phi
_{U}\right|  $ has dimensions of velocity and $\kappa$ is related to the
matter density of the Universe. For critical Universe we know that $\left|
\phi_{U}\right|  /c^{2}=1,$ and thus $E_{0}=mc^{2}.$

\subsubsection{The classical Equivalence Principle}

It is appropriate to make a comment on the Equivalence Principle (EP) at this
point. In our view, as in the Machian view as expressed by Sciama
\cite{sciama}, the EP is a property of the gravitational interaction with the
Universe. The inertial mass is just the gravitational mass scaled by the local
gravitational potential (in units of $c^{2}$) due to the massive Universe.
Therefore, the ratio of the gravitational mass and the inertial mass is a
property of the Universe and not of the test particle! Thus there is no
surprise in the perfect validity of the classical Equivalence Principle. A
violation of the EP can arise only when there are new long range forces that
have properties different from the gravitational coupling.

\subsection{Lorentz invariance of fundamental equations}

We have seen that the Maxwell's equations are invariant under the frame
transformations in Cosmic Relativity, since we deduced that the velocity of
light is a constant in all frames if it is a fundamental constant in the
cosmic frame. We need to ensure that the same Lorentz invariance applies to a
general interval between two space-time events, and also for relativistic wave
equations. 

The interval between two space-time events in the cosmic rest frame is
\begin{equation}
ds^{2}=c^{2}dt^{2}-(dx^{2}+dy^{2}+dz^{2})
\end{equation}
and in another frame moving with respect to the cosmic frame with velocity $V$
is%
\begin{equation}
ds^{\prime2}=c^{2}dt^{\prime2}-(dx^{\prime2}+dy^{\prime2}+dz^{\prime2})
\end{equation}
We have already shown that when $ds=0,$ for light, the two expressions are
identical since velocity of light cannot depend on the state of motion of the
frame, and Lorentz transformations follow from this. Now we apply the Lorentz
transformations to the finite nonzero intervals to see whether these intervals
are invariant. These are
\begin{align}
x^{\prime} &  =\gamma(x-Vt)\nonumber\\
t^{\prime} &  =\gamma(t-Vx/c^{2})
\end{align}
$\gamma=(1-V^{2}/c^{2})^{-1/2}.$

Since $dy^{\prime}=dy$ and $dz^{\prime}=dz,$ we need to consider only
$ds^{\prime2}=c^{2}dt^{\prime2}-dx^{\prime2}.$ Substituting the Lorentz
transformations we see that $ds^{\prime2}=ds^{2}.$Thus what is good for light
is also good for all space-time intervals.

If $E=mc^{2}$ is the energy, $p=mv$ the momentum and $m$ the mass of a
particle in the cosmic rest frame, the Lorentz transformed quantities are
$E^{\prime}=\gamma mc^{2},$ and $p^{\prime}=\gamma mv.$ Then%
\begin{equation}
E^{\prime2}-p^{\prime2}=E^{2}-p^{2}=m^{2}c^{4}%
\end{equation}
Therefore the Lorentz invariance of the Klein-Gordon equation and the Dirac
equation will be valid in Cosmic Relativity. If these equations are valid in
the cosmic rest frame, they are valid in all frames moving relative to the
cosmic frame. The validity of the form of a physical law in all frames in not
an assumption, unlike in SR. If the physical law is valid in the cosmic rest
frame, \emph{it is shown to be valid} in all frames in arbitrary motion
relative to the cosmic rest frame.

\subsection{Conceptual and philosophical implications}

There has been a significant change, in fact the most profound and far
reaching, in the philosophical view on space and time after Einstein's
relativity theory became understood. The development of Cosmic Relativity and
experimental evidence favouring it will imply a large shift in our world-view.
The new world-view will of course be different from the one existed in pre-SR
days, though Cosmic Relativity brings into focus a preferred frame we call the
cosmic rest frame or the absolute frame. Since there is no ether, and since
the new circumstances arise in acknowledging the gravitational presence of the
Universe, a world-view based on Cosmic Relativity will be different from the
one induced by Lorentzian Relativity or Special Relativity. It is important to
note that the only aspect of the cosmos we have used in deducing a new theory
relativity is its approximate homogeneity and isotropy, and the fact that the
Universe is nearly at critical density. There is no use of any General
Relativity. On the other hand, it is possible to start from Einstein's
equations and the standard Robertson-Walker metric as a solution to it, and
then deduce the results of Cosmic Relativity. There is complete consistency
between the two. It may be noted the all the physical effects we talked about
can be derived rigorously in this manner, though it is sufficient to use a
post-Newtonian approximation to relativistic gravity to deduce these effects.
This implies that denying Cosmic Relativity is equivalent to denying the
applicability of results from General Relativity (or any metric or
scalar-tensor relativistic theory of gravitation) to clocks and scales moving
in a massive Universe.

General Relativity follows from Cosmic Relativity more naturally than from
Special Relativity because all general relativistic solutions have to take
into account of the asymptotic or global properties of space and time. Also,
SR is strictly applicable only to inertial observers, whereas Cosmic
Relativity is applicable for general motion. In fact, Einstein himself had
commented on the fact that General Relativity somehow brings back the
possibility of an abstract ether \cite{ein-ether}, without the mechanical
properties that were assigned to it in the pre-SR days.

There is also the question whether there should be a change in our attitude
towards quantizing gravity. Space and time are unobservables, and really has
no meaning in the absence of matter. It is matter that defines, facilitates,
and modifies measurements of spatial and temporal intervals. Therefore,
attempting to quantize space and time (and perhaps worse, space-time as a
single entity) as if they have a fundamental physical existence is perhaps
meaningless in the Machian sense.

I will comment on these aspects in detail in a separate paper. At present it
suffices to mention that everything we know in General Relativity is
consistent with Cosmic Relativity, and the harmony between the two is even
better than in the case of General Relativity and Special Relativity.

Our approach to the theory of relativity answers many fundamental questions
asked in the context of relativity of motion. Already one can see in Sciama's
demonstration \cite{sciama} of inertial reaction arising from the
gravitational interaction of the Universe a synthesis of Newton's first and
second laws, and their relation to his law of gravity \cite{unni-mach}. These
connections are completed by Cosmic Relativity. We can now answer some of the
doubts raised by Julian Barbour in his book, ``Absolute or Relative Motion?:
Discovery of Dynamics'' on such connections \cite{barbour}. Cosmic Relativity
strengthens these connections further, and fully extends Machian ideas to all
of physics by demonstrating that relativistic modifications of spatial and
temporal intervals, as well as several important effects specific to quantum
systems, are the results of the gravitational interaction of the Universe with
the local physical system. The greatest surprise is the revelation that
observed relativistic effects on clocks and scales in flat space-time
themselves are fully Machian.

The construction of Cosmic Relativity also answers the important question
whether there is a second Mach's principle for time \cite{barbour2}. All
modifications of temporal intervals, and even the relativity of simultaneity
arising from the limitation of the maximum velocity possible for any entity in
this Universe are indeed gravitational effects of the Universe. I did not
stress on these aspects in this paper since most of the physicists think that
ideas of Machian flavour are philosophical in nature, without relevance to
physics and its calculations. This is a severe misunderstanding I will attempt
to clarify in a separate paper, addressed to both physicists and philosophers.

\medskip

\noindent\textbf{Acknowledgements}: I have benefitted from encouragement,
suggestions and criticisms by several colleagues and friends in building up
these ideas from considerations of our inseparable links with the Universe.
Comments on the preliminary versions of the paper as well as on the talks
presenting these ideas since December 2003 have helped considerably in
sharpening the ideas, and the presentation. I would like to particularly thank
C. V. K. Baba, A. P. Balachandran, G. T. Gillies, N. Kumar, M. B. Kurup,
Martine Armand, D. Narasimha, R. Nityananda, Anna Nobili, G. Rajasekharan, V.
Radhakrishnan, Joseph Samuel, Venzo de Sabbata, Sundar Sarukkai, Sukant Saran,
Vikram Soni, B. V. Sreekantan, E. C. G. Sudarshan, D. Suresh, A. R. Usha Devi,
and Matt Walhout.

\end{document}